%Paper: hep-th/9410193
%From: Julio Abad <julio@marte.unizar.es>
%Date: Wed, 26 Oct 94 13:58:57 +0100

\input phyzzx
\def \sign{ \mathop{ \rm sign}\nolimits}

{\obeylines
\hfill { DFTUZ /94/21}
\hfill { Sep. 1994 }
\hfill { Hep-Th/mmyynnn }  }
\vskip 40pt plus 2pt minus2pt

\title{INTEGRABLE SPIN CHAINS ASSOCIATED TO $\widehat{sl_{q} \left( n \right)}$
and $\widehat{sl_{p,q} \left( n \right)}$}
\author{J. Abad and M. Rios}
\address{Departamento de F\'{\i}sica Te\'{o}rica, Facultad de Ciencias,
Universidad de Zaragoza, 50009 Zaragoza, Spain}
\date{\today}
\bigskip
\bigskip
\bigskip
\bigskip
\abstract

The Hoft structure of the central extension of the $U_q \left(
\widehat{sl\left( n \right) }\right)$ algebra is considered. The intertwine
matrix induces new integrable spin chain models. We show the relation of these
models and the biparametric spin chain $\widehat{sl_{p,q} \left( n \right)}$
models. The cases $n=2$ are $n=3$ are discussed and for $n=2$ we obtain the
model of Dasgupta and Chowdhury . The case $n=3$ is solved with nested Bethe
ansatz method and it is showed the dependence of the Bethe equations in the
second parameter introduced
\vfill
\eject

\chapter {Introduction}
The search for integrable models is an important problem and has deserved great
attention over the two last decades. The isotropic and anisotropic spin chains
of  Heisenberg occupy a central position in such studies. The mathematical
structure arising in these relatively simple models is astonishingly rich. The
Yan Baxter equation (YBE), based on the original treatment of Baxter and the
quantum inverse formulation due to Faddeev and collaborators $[1]$ are the key
to find new solvable models.

The quantum groups $[2]$ constitute an elegant formalism to obtain, in a
consistent mathematical way, objects fulfilling the YBE and therefor, providing
new solvable spin chains $[3]$.

In this paper, we find a set of integrable models by considering a central
extension of the algebra $ U_{q} \left(\widehat{sl\left( n \right)}\right)$,
that obviously introduce a suitable definition of the coproduct on its Hopf
algebra. The models so obtained, are related with the models derived from the
coloured braid group representations $[4]$ and they are the two parameter
deformed quantum groups ${\widehat{sl_{p,q} \left( n \right)}}$ $[5,6]$.

The present paper is organized as follows. In the next section we develop the
formalism and show the relations with other models in the cases $ n=2$ and $3$.

In third section, the model with $n=3$ is solved by the nested Bethe ansatz
(NBA) method. The Bethe equations obtained, show the dependence on the second
parameter that will introduced a new degree of freedom in  its solutions
compared with the obtained with ${\widehat{sl_{p} \left( n \right)}}$ $[7,8]$.

\chapter {Formulation}
Consider  $ U_{q} \left(\widehat{sl\left( n \right)}\right)$ the universal
covering of the affine algebra and let its generators be
$\left\{  {E_i, F_i, H_i} \right\}_{i=0}^{n-1}$. This algebra has a central
extension $\left\{ Z \right\} $ whose elements are a multiples of the identity,
$Z=\lambda  I$, $ \lambda$ being  a parameter. Then, if $q$ is not a root of
the unity, the elements of the fundamental representation will be characterized
by two parameters, $ \lambda$ and the affine parameter of the algebra $x$
$[9]$.

The $ U_{q} \left(\widehat{sl\left( n \right)}\right)$ with the central
extension has a Hopf algebra. The coproduct is not uniquely determinate, then
we can defined  one coproduct $\triangle$,
 $$\eqalignno{
&\triangle (E_i)=Z K_i \otimes E_i + E_i \otimes K_i^{-1},\qquad
i=1,\cdots,n-1,&(2.1a)\cr
&\triangle (F_i)=Z^{-1} k_i \otimes F_i + F_i \otimes K_i^{-1}, \qquad
i=1,\cdots,n-1,&(2.1b)\cr
&\triangle (E_0)=Z^{-(n-1)} K_0 \otimes E_0+ E_0 \otimes K_0^{-1},&(2.1c)\cr
&\triangle (F_0)=Z^{n-1} K_0 \otimes F_0 + F_0 \otimes K_0^{-1},&(2.1d)\cr
&\triangle (K_i^{\pm 1})=K_i ^{\pm 1}\otimes K_i ^{\pm 1},\qquad
i=0,\cdots,n-1, &(2.1e)\cr
&\triangle (Z)=Z \otimes Z,&(2.1f)\cr
}
$$
and its coproduct symmetric $\triangle^\prime$,
 $$\eqalignno{
&\triangle^\prime (E_i)=K_i^{-1} \otimes E_i + E_i \otimes Z K_i,&(2.2a)\cr
&\triangle^\prime(F_i)= K_i ^{-1}\otimes F_i + F_i \otimes Z^{-1}K_i^{-1},
&(2.2b)\cr
&\triangle^\prime (E_0)= K_0^{-1} \otimes E_0+ E_0 \otimes
Z^{-(n-1)}K_0^{-1},&(2.2c)\cr
&\triangle^\prime (F_0)= K_0^{-1} \otimes F_0 + F_0 \otimes
Z^{n-1}K_0^{-1},&(2.2d)\cr
&\triangle^\prime (K_i^{\pm 1})=K_i ^{\pm 1}\otimes K_i ^{\pm 1}, &(2.2e)\cr
&\triangle^\prime (Z)=Z \otimes Z.&(2.2f)\cr
}
$$
Both coproducts must be related by a transformation matrix $R_{1,2} $ such that
$$
R_{1,2}  \cdot \triangle_{(x,\lambda)\otimes(y,\mu)}(a) =
\triangle_{(x,\lambda)\otimes(y,\mu)}^{\prime} (a)  \cdot R_{1,2} ,\qquad
\forall a \in U_q (\widehat{sl(n)} ),\eqno (2.3)
$$
where $R$ verifies the Yang-Baxter equation
$$ R_{1,2} \cdot R_{1,3}  \cdot  R_{2,3} =  R_{2,3}  \cdot R_{1,3}  \cdot
R_{1,2}. \eqno(2.4)$$
that is the key to build an integrable model  $[10]$.

The solution, that we have found to eq. (2.3), is the form
$$
\eqalign{
R_{1,2} (x,y,\lambda,\mu)=&
\left( y^n-q^2 x^n \right) \sum_{i=1}^{n}{\lambda^{i-1}\mu^{n-i}e_{i,i}\otimes
e_{i,i} } +
 q ( y^n - x^n ) \sum_{i,j=1 \atop i \neq
j}^{n}{\lambda^{j-1}\mu^{n-i}e_{i,i}\otimes e_{j,j}}  \cr
+(1-q^2)  &\sum_{i,j=1 \atop i < j}^{n}{( \lambda^{i-1}\mu^{n-i} x^{j-i} y^{i-j
}y^n e_{i,j}\otimes e_{j,i}+
\lambda^{j-1}\mu^{n-j} x^{i-j} y^{j-i} x^n e_{j,i}\otimes e_{i,j})} .\cr} \eqno
(2.5)
$$
Associated to every solution of YBE acting on the spaces $C_1^n \otimes C_2^n$,
we can find a solvable model  $[8]$. So, we introduce an one-dimensional
lattice with a vector space $V_r \equiv C^n$ in every site. Now, we define an
operator per site equal to $R_{1,2}\left( x,y,\lambda,\lambda_{0} \right)$
where the first space $C_1^n$ is an auxiliary space and the second space is the
$V_r$. We call this operator
$$
\eqalign{
L_r (u, \lambda, \lambda_{0}) =& \sinh{( {n \over 2} u +\gamma) }
 \sum_{i=1}^{n}{ \left( {\lambda \over  \lambda_{0}} \right)^i e_{i,i}\otimes
e_{i,i} ^r}+
\sinh{({n \over 2}  u)} \sum_{i,j=1 \atop i \neq j}^{n}{  \left( {
{\lambda}^{j}\over {\lambda}_{0}^{i}} \right)
e_{i,i}\otimes e_{j,j}^r }    \cr &+
\sinh{(\gamma)}  \sum_{i,j=1 \atop i \neq j}^{n} {    exp{[(i-j-{n \over 2}
\sign {(i-j)}) u]}
  \left( {\lambda \over  \lambda_{0}} \right)^i e_{i,j}\otimes e_{j,i}^r
}.\cr} \eqno(2.6)
$$
where we have made the substitutions
$$
{y \over x} = \exp{(u)} ,\qquad q = \exp{(-\gamma)}. \eqno (2.7)
$$
The YBE can be written now as
$$
R(u-v,\lambda,\mu) \cdot \left( L_r (u,\lambda,\lambda_{0}) \otimes L_r
(v,\mu,\lambda_{0}) \right) = \left( L_r (v,\mu,\lambda_{0}) \otimes L_r
(u,\lambda,\lambda_{0})  \right)\cdot R( u-v ,\lambda,\mu), \eqno (2.8)
$$
where the $\otimes$ product is in the site space and the $\cdot$ product is in
the $A \otimes A$ tensorial space. The operator $R$ in (2.8) is obtained from
$R_{1,2}$ in $(2.5)$  by interchanging the indices $j$ and $m$ in every product
 $e_{i,j}\otimes e_{l,m}$ and the same substitution on its accompanying
coefficient. So
$$
\eqalign{
R (u,\lambda,\mu) = & \sinh{( {n \over 2} u +\gamma) }  \sum_{i=1}^{n}{ \left(
{\lambda \over  \mu} \right)^i e_{i,i}\otimes e_{i,i} }+
\sinh{({n \over 2}  u)} \sum_{i,j=1 \atop i \neq j}^{n}{  \left( {
{\lambda}^{i}\over \mu^{j}} \right)e_{i,j}\otimes e_{j,i} }   \cr & +
\sinh{(\gamma)}  \sum_{i,j=1 \atop i \neq j}^{n} {    exp{[(j-i-{n \over 2}
\sign {(j-i)}) u]}
 \left( {\lambda \over  \mu} \right)^j e_{i,i}\otimes e_{j,j}   },\cr}
\eqno(2.9)
$$
With the local operator $L_r$, we build the monodromy matrix
$T(u,\lambda,\lambda_0)$ defined on the auxiliary space $A$, whose components
are operators on the configuration space, i.e. the tensorial product of the
site spaces of the lattice
$$
T(u,\lambda,\lambda_0 ) \equiv T(u,\lambda) = L_N (u,\lambda,\lambda_0) \cdot
L_{(N-1)} \cdots L_1 (u,\lambda,\lambda_0), \eqno(2.10)
$$
where the $\cdot$ product is understood as before in the auxiliary space.

The monodromy matrix $T$ enjoys most of the properties of $L_r$. The most
important is a YBE similar to (2.7)
$$
R(u-v,\lambda,\mu) \cdot \left(T (u,\lambda,\lambda_{0}) \otimes T
(v,\mu,\lambda_{0}) \right) = \left( T (v,\mu,\lambda_{0}) \otimes
T(u,\lambda,\lambda_{0})  \right)\cdot R( u-v ,\lambda,\mu). \eqno (2.11)
$$
A consequence of (2.11) is the existence of a commuting family of transfer
matrices $F$, given by the expression
$$
F(u,\lambda,\mu)=trace_{aux} T(u,\lambda,\mu),\eqno (2.12)
$$
for which
$$
\left[ F(u,\lambda,\lambda_0), F(v,\mu,\lambda_0) \right]=0, \eqno (2.13)
$$
as can be proved by taking the trace of  $(2.11)$. This implies that
differentiating $F$ respect his parameters for certain values of them, we
obtain a set of commuting operators. In particular the hamiltonian, a two next
neighbors interacting operator, is related to the first logarithmic derivative
of $F$. So, we can take the hamiltonian as
$$
H = {2 \over n} \sinh{\gamma} {\partial \over \partial u}{ \ln (F(u))} \big |
_{u=0 \atop \lambda=\lambda_0} - {N \over n} \cosh{\gamma},  \eqno(2.14)
$$

The derivative respect to $\lambda$
$$
Q=- \lambda_0 {\partial \over \partial \lambda}{ \ln (F)} \big | _{u=0 \atop
\lambda=\lambda_0} +\left( {n+1 \over 2} \right) N, \eqno(2.15)
$$
will be a conserved charge. These operators can be expressed as
$$
\eqalignno{
&H=\sum_{r=1}^{N-1}{h_{r,r+1} }, &(2.16a) \cr
&Q=\sum_{r=1}^{N-1}{k_{r,r+1}}, &(2.16b) \cr
}$$
with
$$
\eqalign{
h_{r,r+1} = & {n-1 \over n} \cosh{(\gamma)}    \sum_{i=1}^{n}{ e_{i,i}^{r}
\otimes e_{i,i}^{r+1} } +\sum_{i,j=1 \atop i \neq j}^{n}{
\lambda_0^{(i-j)}e_{i,j}^{r} \otimes e_{j,i}^{r+1}} \cr & +
\sum_{i,j=1 \atop i \neq j}^{n}{ \Bigl(  \bigl( {2 (j-i) \over n} - \sign
(j-i)\bigr) \sinh{( \gamma )} - {\cosh{( \gamma )} \over n}\Bigr)  e_{i,i}^{r}
\otimes e_{j,j}^{r+1} } .\cr}
\eqno (2.17)
$$
and
$$
\eqalign{
k_{r,r+1} = & -    \sum_{i=1}^{n}{ i e_{i,i}^{r} \otimes e_{i,i}^{r+1} }
+\sum_{i,j=1 \atop i \neq j}^{n}{  j e_{i,j}^{r} \otimes e_{j,i}^{r+1}}  +
{n+1 \over 2} \sum_{i,j=1 \atop i \neq j}^{n}{  e_{i,i}^{r} \otimes
e_{j,j}^{r+1} } \cr &
=\sum_{i,j=1 \atop i \neq j}^{n}{ \left( {n+1 \over 2} -j \right) e_{i,i}^{r}
\otimes e_{j,j}^{r+1} } = I^r \otimes S_z^{\left( r+1 \right)}.\cr}
\eqno (2.18)
$$
The  last expression shows that $k$ is a local operator, the third component of
the spin $su(2)$. The operator $Q$, in view of $(2.16)$, is the sum of the
these components and it is a conserved quantity.

If we specify for $n=2$ and $\lambda=\lambda_0=\exp{i \delta}$, we obtain
$$
H_{sl\left( 2 \right)}={1 \over 2}
\sum_{i=1}^{N}
{\left( \cos{\delta} \left( \sigma _x^i \sigma_x^{i+1 }+\sigma _y^i  \sigma_y^{
i+1 } \right)+
\cosh{\gamma}  \sigma _z^i \sigma_z^{  i+1 }+
 \sin{\delta}  \left( \sigma _y^i \sigma_x^{ i+1 } - \sigma _x^i \sigma_y^{ i+1
}
\right)\right)
}, \eqno(2.19)
$$
where $\sigma$ are the Pauli matrices.

For $n=3$ the hamiltonian obtained is
$$
\eqalign{
H_{sl(3)} = {1 \over 2}  \sum_{i=1}^{N}
  &\Bigl( \cos{\delta} ( \lambda_1^i  \lambda_1^{i+1} +\lambda_2^i
\lambda_2^{i+1} +
\lambda_6^i  \lambda_6^{i+1} +\lambda_7^i  \lambda_7^{i+1} )           \cr
& + \sin{\delta}  ( \lambda_2^i  \lambda_1^{i+1} -\lambda_1^i  \lambda_2^{i+1}
+\lambda_7^i  \lambda_6^{i+1} -\lambda_6^i  \lambda_7^{i+1} )           \cr
&+\cos{(2 \delta)} \left( \lambda_4^i \lambda_4^{i+1}+\lambda_5^i
\lambda_5^{i+1}\right)+
\sin{(2 \delta)} \left( \lambda_5^i \lambda_4^{i+1}-\lambda_4^i \lambda_5^{i+1}
\right) \cr
&+\cosh{\gamma} ( \lambda_3^i \lambda_3^{i+1}+\lambda_8^i \lambda_8^{i+1})+
{ \sinh{\gamma}\over \sqrt{3} } ( \lambda_8^i \lambda_3^{i+1}-\lambda_3^i
\lambda_8^{i+1})
 \Bigr),}
\eqno (2.20)$$
where we have used the same substitutions that before and $\lambda$ are the
Gell-Mann matrices.

For $\delta=0$ these hamiltonians correspond to the XXZ models and their
generalizations to  $sl(n)$ $[8]$.

A more specific model is obtained if we do the substitutions
$$
\exp{(-\gamma)}=\sqrt{ p q},\qquad  \exp{(i \delta)}=\sqrt{ {p \over q}}
.\eqno(2.21)
$$
we find in this way the models obtained from $SU_{p,q}(n)$. Since that for
$n=2$ we obtain the model of Dasgupta and Chowdhury $[5]$, there must be  a
relation between $U_r (sl(2))$ and
$U_{p,q} (sl(2))$. In fact, the set $\{ e, f , k^{\pm 1}\equiv r^{\pm {h \over
2}} \}$ of generators of $U_r (sl(2))$ and the set $\{ \tilde{e}, \tilde{f}
,q^{\pm {h \over 2}}, p^{\pm {h \over 2}} \}$ of generators of $U_{p,q}
(sl(2))$ are $\{ \tilde{e}, \tilde{f} ,q^{\pm {h \over 2}}, p^{\pm {h \over 2}}
\}$, that verify respectively the equations
$$
[e, f]= {{r^h -r^{-h}} \over {r-r^{-1}}},\qquad [\tilde{e}, \tilde{f}]_{({p
\over q})^{{1 \over 2}}}\equiv \tilde{e} \tilde{f} - p q^{-1}\tilde{f}
\tilde{e} = {{q^h -p^{-h}} \over {q-p^{-1}}}, \eqno (2.22)
$$
are related to each other by
$$
\tilde{e}=\left( { q\over p}\right)^{{h \over 4}} e,\qquad \tilde{f}=\left(  {
q\over p}\right)^{{h \over 4}} f .\eqno (2.23)
$$

In this sense, the models we derived from the quantum group  $sl(n)$ with
center include the model derived by Dasgupta and Chowdhury.

\chapter{Bethe solutions in the $n=3$ case}

The usual method to solve these models is the algebraic Bethe ansatz proposed
by Faddeev and his collaborators $[1]$. For a model with site space of $n$
components, the method, know as nested Bethe ansatz $[7,8]$, is developed in
$(n-1)$ steps,every one similar to the Bethe ansatz. In this section we are
going to solve the case $n=3$ and we will show the main features of the model.
The generalization to higher values of $n$ of the conserved magnitudes will
follow immediately.

We start by specifying the monodromy operator (2.10) as
$$
T(u,\lambda)=T(u,\lambda,\lambda_0)=\pmatrix{
A(u,\lambda)& {B}_{2} (u,\lambda) & {B}_{3}(u,\lambda) \cr
{C}_{2} (u,\lambda) & {D}_{2 2} (u,\lambda) & {D}_{2 3}(u,\lambda) \cr
{C}_{3} (u,\lambda) & {D}_{3 2} (u,\lambda) & {D}_{3 3}(u,\lambda) \cr
}
\eqno (3.1)
$$

The components are operators in the configuration space of the lattice.
Considering
$$
B(u,\lambda)=\left (\matrix{
{B}_{2} (u,\lambda)  & {B}_{3} (u,\lambda) \cr
}\right ), \qquad
D(u,\lambda)=\pmatrix{
{D}_{2 2} (u,\lambda)  & {D}_{2 3} (u,\lambda)  \cr
{D}_{3 2} (u,\lambda)  & {D}_{3 3} (u,\lambda) \cr
} ,\eqno(3.2)
$$
the YBE (2.8) gives the relations
$$
\eqalignno{
&B(u,\lambda) \otimes B(v,\mu)= \bigl[ B(v,\mu)\otimes B(u,\lambda) \bigr]\cdot
 {R}^{(2)} (u-v,\lambda,\mu), & (3.3a) \cr
&A(u,\lambda) B(v,\mu)=g(v-u) B(v,\mu)  A(u,\lambda) s(\lambda)
\cr &  \qquad \qquad \qquad \qquad  -B(u ,\lambda)) A(v,\mu) \cdot
{\tilde{r}}^{(2)}(v-u) s(\lambda),& (3.3b) \cr
&D(u,\lambda) \otimes B(v,\mu)=g(u-v) B(v,\mu) \otimes s(\mu)(D(u,\lambda)
\cdot{R}^{(2)} (u-v,\lambda,\mu) ) \cr &
\qquad \qquad \qquad \qquad -B(u,\lambda) \otimes s(\lambda) ({r}^{(2)}(u-v)
\cdot D(v,\mu) ) , & (3.3c) \cr
}
$$
$g$ and $h_{\pm}$ being  the functions
$$
g(\theta)={\sinh({n \over 2} \theta+\gamma) \over\ sinh{({n \over 2}\theta})}
,\qquad
h_{\pm}={\sinh{(\gamma)}  e^{\pm{\theta \over 2}} \over \sinh{({n \over
2}\theta)}},
\eqno(3.4)
$$
and the matrices
$$
s(x)=\pmatrix{
x & 0 \cr
0 & x^2 \cr
},\qquad
{r}^{(2)}(\theta)=\pmatrix{
{h}_{-}(\theta)& 0 \cr
0 & {h}_{+}(\theta)\cr
},\qquad
{\tilde{r}}^{(2)}(u)=
\pmatrix{
{h}_{+}(\theta)& 0 \cr
0 & {h}_{-} (\theta)\cr
},
$$
$$
R^{(2)}(u,\lambda,\mu)=
\left (\matrix{
{\lambda \over \mu} & 0 & 0 & 0 \cr
0 & { h_{-}(u)\over g(u)} {\lambda^2 \over \mu^2} & { 1\over g(u)} {\lambda
\over \mu^2}  & 0 \cr
0 & { 1\over g(u)} {\lambda^2 \over \mu} & { h_{+}(u)\over g(u)} {\lambda \over
\mu} & 0 \cr
0 & 0 & 0 & {\lambda^2 \over \mu^2} \cr
}\right ).\eqno(3.5)
$$

The state
$$
 \parallel 1>={
\left (\matrix{
1\cr
0\cr
0\cr
}\right )}_{N} \otimes \cdots \otimes
{
\left (\matrix{
1\cr
0\cr
0\cr
}\right )}_{1} .\eqno(3.6)
$$
is an eigenstate of the $A$ and $D_{i,j}$ components of $T$, i.e.
$$
\eqalignno{
A(u,\lambda) \parallel 1> &={[\sinh({3 \over 2}u+\gamma) {\lambda \over
\lambda_0} ]}^N \parallel 1>={[a(u,\lambda)]}^N \parallel 1> &(3.7a )\cr
D_{i,j}(u,\lambda) \parallel 1> &={[\sinh({3 \over 2}u+\gamma) {\lambda \over
\lambda_0^i} ]}^N \delta_{i,j} \parallel 1>={[d_{i,j}(u,\lambda)]}^N \parallel
1> &(3.7b )\cr
}
$$
Now, with the help of the relations (3.3), we look for solutions of
the equation
$$
F(u,\lambda) \Psi
_{\lambda_0}({\mu}_{1},\cdots,{\mu}_{r})=\Lambda(u,\lambda,\lambda_0,
{\mu}_{1},\cdots,{\mu}_{r}) \Psi _{\lambda_0}({\mu}_{1},\cdots,{\mu}_{r}),\eqno
(3.8)
$$
of the form
$$
\Psi_{\lambda_0}(\vec{\mu})=\Psi_{\lambda_0}
({\mu}_{1},\cdots,{\mu}_{r})={X}_{{i}_{1},\cdots,{i}_{r}}
{B}_{{i}_{1}}({\mu}_{1})\otimes \cdots \otimes{B}_{{i}_{r}}({\mu}_{r})
\parallel 1>, \eqno (3.9)
$$
To begin with, since $\parallel 1>$ is eigenvector of  $A(u)$ and ${D}_{i,i}$,
we apply these operators on $\Psi$ and, by using the commutation relations
(3.3), we push the operators $A$ or $D_{i,j}$ through the $B$ to the right .
When either $A$ or $D$ reaches $\parallel 1> $ they reproduce this vector
again. Since the commutation relations have two terms, this procedure generates
a lot of terms. Some of them have the same order of the arguments in the $B$
product; we call them wanted terms. The others have some
$B({\mu}_{j}\lambda_0)$ replaced by $B(u,\lambda_0)$ and we call them unwanted
terms.

When we apply  $F= A+D_{2,2} +D_{3,3}$  to $ \Psi
({\mu}_{1},\cdots,{\mu}_{r})$, we collect the unwanted terms and require them
to have a vanishing sum. This condition gives us a set of equations for the
parameters. The sum of the wanted terms will be required to be proportional to
$\Psi$, providing us with the second part of equation (3.8).

So, the application of  $F(u,\lambda)$ on $\Psi_{\lambda_0}(\vec{\mu})$ gives
the wanted term

$$
\eqalign {
&\biggl[ a(u,\lambda)^N \prod_{j=1}^{r}{g({\mu}_{j} - u)}
{B}_{{j}_{1}}({\mu}_{1},\lambda_0)\otimes \cdots \otimes
{B}_{{j}_{r}}({\mu}_{r},\lambda_0)  S^{(r)}(\lambda) {X}_{{j}_{1}, \cdots,
{j}_{r}} \cr
&+
\prod_{j=1}^{r}{g(u-{\mu}_{j} )}  {B}_{{j}_{1}}({\mu}_{1},\lambda_0)\otimes
\cdots \otimes  {B}_{{j}_{r}}({\mu}_{r},\lambda_0)  F_{(2)}^{(r)}(u,
\vec{\mu},\lambda,\lambda_0) {X}_{{j}_{1}, \cdots, {j}_{r}}\biggr] \parallel
1>.  \cr} \eqno(3.10)
$$
being
$$
\eqalignno{
&S^{(r)} (\lambda)= \underbrace {s(\lambda) \otimes\cdots \otimes
s(\lambda)}_{r-times} &(3.11a)\cr
&F_{(2)}^{(r)}(u,\vec{\mu},\lambda,\lambda_0)=\lambda_0^r d_{2 2}^N (u,\lambda)
A^{(2)}(u,\vec{\mu},\lambda)+\lambda_0^{2r} d_{3 3}^N (u,\lambda)
D^{(2)}(u,\vec{\mu},\lambda)&(3.11b)}
$$

Now, we impose  the cancelation of the unwanted terms.
The operators $A^{(2)}$ and $D^{(2)}$ are components of
$$
\eqalign {
T_{r}^{\left( 2
\right)}(u,\vec{\mu},\lambda,\lambda_0)_{j,{j}_{1},\cdots,{j}_{r}}^{i,{i}_{1},\cdots,{i}_{r}}&=
{{R}^{\left( 2 \right)  }}_{  {j}_{r},{a}_{r}}^{a_{r-1},{i}_{r}}
(u-{\mu}_{r},\lambda,\lambda_0)\cdots \cr
\cdots &
{{R}^{\left( 2 \right)  }}_{  {j}_{2},{a}_{2}}^{{a}_{1},{i}_{2}}
(u-{\mu}_{2},,\lambda,\lambda_0)
{{R}^{\left( 2 \right)  }}_{  {j}_{1},a_1}^{ j,{i}_{1}}
(u-{\mu}_{1},\lambda,\lambda_0),
 \cr}\eqno(3.12)
$$
a $2\times2$ matrix in the components acting on the second and third components
of the auxiliary space, that can be written
$$T_{r}^{\left( 2 \right)}(u,\vec{\mu},\lambda,\lambda_0))=
\pmatrix{
A_{r}^{\left( 2 \right)}(u,\vec{\mu},\lambda,\lambda_0)) & B_{r}^{\left( 2
\right)}(u,\vec{\mu},\lambda,\lambda_0))\cr
C_{r}^{\left( 2 \right)}(u,\vec{\mu},\lambda,\lambda_0)) & D_{r}^{\left( 2
\right)}(u,\vec{\mu},\lambda,\lambda_0)) \cr
},
\eqno(3.13)
$$
and its components are operators on the configuration space.

Then, in order to $\Psi $ be solution of $(3.8)$, we must require
$$\eqalignno{
&S^{(r)}(\lambda) X = \omega_r (\lambda) X &(3.14a) \cr
&F_{(r)}^{\left( 2 \right)}(u,\vec{\mu},\lambda,\lambda_0))
X=\Lambda_{(r)}^{\left( 2 \right)}(u,\vec{\mu},\lambda,\lambda_0)) X &(3.14b)
\cr}
$$

The cancelation of the unwanted terms impose the set of equations
$$
[a(\mu_k,\lambda_0)]^N \omega_{r-1}\left( \lambda_0 \right)=\prod_{j \ne k
\atop j=1}^{r}{{g\left( {\mu}_{k}-{\mu}_{j} \right) \over g\left(
{\mu}_{j}-{\mu}_{k}\right) }\Lambda_{(r-1)}^{\left( 2
\right)}(\mu_k,\vec{\mu},\lambda,\lambda_0) },\qquad k=1,\cdots,r\eqno(3.15)
$$

The second step is to diagonalize the (3.11b) equation. We apply the same
method as in the first step with one unit lower. So, the operator
$T_{r}^{\left( 2 \right)}$ verifies the YBE
$$ \eqalignno {
R^{(2)}\left( u-v,\lambda,\mu \right) \cdot &\left(    T_{r}^{\left( 2
\right)}(u,\vec{\mu},\lambda,\lambda_0) \otimes
T_{r}^{\left( 2 \right)}(v,\vec{\mu},\mu,\lambda_0\right)= &\cr
&=\left(    T_{r}^{\left( 2 \right)}(v,\vec{\mu},\mu,\lambda_0\) \otimes
T_{r}^{\left( 2 \right)}(u,\vec{\mu},\lambda,\lambda_0))\right)\cdot
R^{(2)}\left( u-v ,\lambda,\mu\right), &(3.16) \cr
}
$$
that gives the relations
$$  \eqalignno {
& {B}^{(2) }(u,\lambda) \cdot  {B}^{(2) }(v,\mu) ={ \lambda\over\mu} {B}^{(2)
}(v,\mu) \cdot  {B}^{(2) }(u,\lambda). &(3.17a)\cr
&{A}^{(2)}(u,\lambda) \cdot {B}^{(2)}(v,\mu)= \lambda g(v-u) {B}^{(2)}(v,\mu)
\cdot {A}^{(2)}(u,\lambda)- &\cr
& \qquad \qquad \qquad \qquad  \qquad \qquad \qquad \qquad- \lambda
{h}_{+}(v-u) {B}^{(2)}(u,\lambda) \cdot {A}^{(2)}(v,\mu), &  (3.17b) \cr
&{D}^{(2)}(u,\lambda) \cdot {B}^{(2)}(v,\mu)= \lambda g(u-v) {B}^{(2)}(v,\mu)
\cdot {D}^{(2)}(u,\lambda)-&\cr
& \qquad \qquad \qquad \qquad  \qquad \qquad \qquad \qquad -\lambda
{h}_{-}(u-v) {B}^{(2)}(u,\lambda) \cdot {v}^{(2)}(v,\mu), &  (3.17c) \cr
}
$$

Now we take the state
$$
 \parallel 1>^{(2)}=
\left (\matrix{
1\cr
0\cr
}\right )_1 \otimes \cdots \otimes
\left (\matrix{
1\cr
0\cr
}\right )_r
\eqno(3.18)
$$
that is a eigenstate of $F^{(2)}$, and we look for eigenstates of the form
$$X={\Psi}^{(2)}={B}^{(2)}(\rho_1,\vec{\mu} ,\lambda_0) \cdots
{B}^{(2)}({\rho}_{s},\vec{\mu} ,,\lambda_0) {\parallel 1>}^{(2)}, \eqno (3.19)
$$
that introduce the dependence of the eigenvalues on a new set of parameters
$\left\{ {\lambda}_{i} \right\}_{i=1}^{s}$.

Following the same procedures as in the first step, but now in two
dimensions,then we find the eigenvalues of $F^{(2)}$ and the conditions that
must verify the set of parameters $\left\{ {\rho}_{i} \right\}_{i=1}^{s}$. Then
we obtain finally the eigenvalues of $F$ and $F^{(2)}$
$$\eqalignno{
\Lambda (u,\vec{\mu} ,\vec{\rho},\lambda,\lambda_0)=&[\sinh{({3 \over
2}u+\gamma)}]^N {{\lambda}^{N+r+s} \over
{\lambda_0}^N}\prod_{j=1}^{r}{g({\mu}_{j}-u)}
&\cr & \qquad
+\prod_{j=1}^{r}{g(u-{\mu}_{j})}
 \Lambda_{(2)}^{r)}(u,\vec{\mu} ,\vec{\rho},\lambda,\lambda_0),
& (3.20a)\cr
\Lambda_{(r)}^{(2)}(u,\vec{\mu} ,\vec{\lambda},\lambda,\lambda_0 )=&
[\sinh{({3 \over 2}u)}]^N {{\lambda}^{N+r+s} \over {\lambda_0}^{2N}}
\Bigl( \prod_{i=1}^{s}{g({\rho}_{i}-u)}
& \cr&\qquad
+{1 \over {\lambda_0}^{N}}\prod_{i=1}^{s}{g(u-{\rho}_{i})}
\prod_{j=1}^{r}{{1 \over g(u-{\mu}_{j})}} \Bigr), &(3.20b)\cr
}
$$
and the parameters$\left\{ {\mu}_{i} \right\}_{i=1}^{r}$ and $\left\{
{\rho}_{i} \right\}_{i=1}^{s}$
solutions of the equations given by the cancelation of the unwanted terms
$$\eqalignno{
&\left( g( {\mu}_{k})\right)^{N} {{\lambda}_{0}}^N=
\prod_{j=1\atop j \ne k}^{r}{{g({\mu}_{k}-{\mu}_{j})\over
g({\mu}_{j}-{\mu}_{k})}}
\prod_{i=1}^{s}{g({\rho}_{i}-{\mu}_{k})}
,  &(3.21a)\cr
&{{\lambda}_{0}}^N\prod_{j=1}^{r}{g({\rho}_{k}-{\mu}_{j})} =
\prod_{i=1\atop i \ne k}^{s}{{g({\rho}_{k}-{\rho}_{i})\over
g({\rho}_{i}-{\rho}_{k})}}. &(3.21b)
\cr}
$$
Every set of solutions for $1 \leq s \leq r\leq N $ of these coupled equations
determines an eigenvalue of $F$.

An analogous set of equations exist for $ \widehat{sl_q (3)}$ model, as you can
see in the references $[7,8]$, the difference between that set and (3.21) is
the factor ${{\lambda}_{0}}^N$ that will modified the solutions for the
parameters  $\left\{ \mu_j \right\}_{j=1}^r$ and $\left\{ \rho_i
\right\}_{i=1}^s$.

The energy spectrum, obtained by applying $(2.14)$ to $\Lambda (u,\vec{\mu}
,\vec{\rho},\lambda,\lambda_0)$, is
$$
E={2 \over 3} N \cosh{(\gamma)} +\sinh{(\gamma)}\sum_{i=1}^{r}{\Bigl({1
\over\tanh{({ 3\over 2}\mu_i} )}-{1 \over\tanh{({ 3\over 2}\mu_i+\gamma})
}\Bigr)}\eqno (3.22)
$$
As we can see in the last expression, the energy depends only on the first set
of  introduced parameters $\left\{ \mu_j \right\}_{j=1}^r$

The second operator defined by $(2.15)$, gives the conserved quantity
$$
q=-\lambda_0 {\partial \over \partial \lambda}{\ln{\Lambda}}\Big | _{u=0 \atop
\lambda=\lambda_0}+2N=N-\left( r+s \right) \eqno (3.23)
$$
which is the third component of a chain of spin $1$ states of a $SU(2)$ group
with $(N-r)$ sites in the state $\vec{e_1}$,  $(r-s)$ sites in $\vec{e_2}$ and
$s$ states in $\vec{e_3}$.

In conclusion, we can say that the introduction of a coproduct with an element
of the center of the algebras, permit to find integrable models with new
parameters. We have show that such models are related with the models coming
from the algebras $ sl_{p,q} (n)$ . In addition, we have found the form
$(2.21)$ to connect the uniparametric models with the multiparametric
deformations.

{\bf Acknowledgements}

We would like to thank G. Sierra for very useful discussions and
Professor J. Sesma for the careful reading of the manuscript. This work was
partially supported by the Direcci\'{o}n General de Investigaci\'{o}n
Cient\'{\i}fica y T\'{e}cnica, Grant No PB93-0302
\vfill
\eject

%
%    Aqui empiezan las referencias
%
%
\centerline{\bf References}

\item{[1]}{L. D. Faddeev, Sov. Sci. Rev. Math. Phys. C1, (1981) 107.}
\item{[2]}{V.G. Drinfeld, Proceedings of the I. C. M. 1986, A.M. Gleason editor
(A.M.S.1987).}
\item{[3]}{A. Berkovich, C. G\'{o}mez and G. Sierra, Int. J. Mod. Phys. B 7,
(1992) 1939. \hfill\break
V. Pasquier and H. Saleur, Nucl. Phys. B 330 (1990), 523.  \hfill\break
C. G\'{o}mez and G. Sierra, Phys. Lett. B 285 (1992), 126.}
\item{[4]}{Y. Akutsu and T. Deguchi, Phys. Rev. Lett. 67 (1991), 777}
\item{[5]}{N. Dasgupta and A. R. Chowdhury, Jour. Phys. A 26 (1993), 5427.}
\item{[6]}{M. Kibler, LYCEN preprint 9338, (1993) (hep-th
9407050).}\hfill\break
{V. Karimipour, Jour. Phys. A 26 (1993), 6277.} \hfill\break
{B. Basu-Mallick, hep-th 9402142, (1994).}
\item{[7]}{H.J. de Vega, Int. J. Mod. Phys. A 4 (1989), 2371.}
\item{[8]}{J. Abad and M.Rios, Univ. de Zaragoza,  preprint DFTUZ 94-11,
(1994).}
\item{[9]}{J. Fuchs, Affine Lie Algebras and Quantum Groups. Cambridge
University Press,\hfill\break
Cambridge (1993).}
\item{[10]}{M. Jimbo, Nankai Lectures on Mathematical Physics 1991, edts C. N.
Yang and M. L. Ge, World Scientific, Singapore.  }

\end
\end{document}

\end{document}